\documentclass[aps,prl,twocolumn,showpacs]{revtex4}
\usepackage{graphicx,color}

\begin{document}

\title{Superconductivity and antiferromagnetism in the two-dimensional\\ 
Hubbard model: a variational study}

\author{D.\ Eichenberger and D.\ Baeriswyl}
\affiliation{Department of Physics, University of Fribourg, 
CH-1700 Fribourg, Switzerland.}

\date{\today}

\begin{abstract}
A variational ground state of the repulsive Hubbard model on a square 
lattice is
investigated numerically for an intermediate coupling strength ($U=8t$) and
for moderate sizes (from $6\times 6$ to $10\times 10$). Our ansatz is clearly
superior to other widely used variational wave functions. The results for
order parameters and correlation functions provide new insight for
the antiferromagnetic state at half filling as well as strong evidence for a 
superconducting phase away from half filling.
\end{abstract}


\maketitle

The Hubbard model plays a key role in the analysis of correlated electron 
systems, and it is widely used for describing quantum antiferromagnetism, 
the Mott metal-insulator transition and, ever since Anderson's suggestion
\cite{Ande1}, superconductivity in the layered cuprates. Several approximate techniques
have been developed to determine the various phases of the two-dimensional
Hubbard model. For very weak coupling, the perturbative Renormalization Group extracts
the dominant instabilities in an unbiased way,
namely antiferromagnetism at half filling and $d$-wave superconductivity 
for moderate doping \cite{Zanc1, Halb1}. Quantum Monte Carlo simulations
have been successful in extracting the antiferromagnetic correlations at 
half filling \cite{Hirs1, Whit1}, but in the presence of holes the numerical
procedure is plagued by the fermionic minus sign problem \cite{Dago1}.
This problem appears to be less severe in dynamic cluster Monte Carlo
simulations, which exhibit a clear tendency towards $d$-wave superconductivity
for intermediate values of $U$ \cite{Maie1}.

Variational techniques address directly the ground state and thus offer an 
alternative to quantum Monte Carlo simulations, which are limited 
to relatively high temperatures. Previous variational wave functions include
mean-field trial states from which configurations with doubly occupied sites
are either completely eliminated (full Gutzwiller projection) 
\cite{Gros1,Yoko1,Para1} or at least partially suppressed \cite{Giam1}.
Recently, more sophisticated wave functions have been proposed, which include,
besides the Gutzwiller projector, non-local operators related to charge and
spin densities \cite{Sore1, Yoko2}. Our own variational wave function is 
based on the idea that for intermediate values of $U$
the best ground state is a compromise between the
conflicting requirements of low potential energy (small double occupancy) and
low kinetic energy (delocalization). It is known that the addition of
an operator involving the kinetic energy yields an order of magnitude
improvement of the ground state energy with respect to a wave function with
a Gutzwiller projector alone \cite{Otsu1}. In this letter, we show that such 
an additional term allows us to draw an appealing picture of the ground 
state, both at half filling and as a 
function of doping (some preliminary results have been published
\cite{Eich1, Baer1}).

In its most simple form, the 2D Hubbard model is composed of two terms,  
$\hat{H}=t\hat{T}+U\hat{D}\, ,$ with 
\begin{equation}
\quad\hat{T}=
-\sum_{\langle i,j\rangle,\sigma}(c^\dag_{i\sigma}c_{j\sigma}
+c^\dag_{j\sigma}c_{i\sigma})\,\  \mbox{and} \ 
\hat{D}=\sum_i n_{i\uparrow}n_{i\downarrow}\, .
\end{equation}
Here $c^\dag_{i\sigma}$ creates an electron at site $i$ with spin $\sigma$, 
the summation is restricted to nearest-neighbor sites and 
$n_{i\sigma}=c^\dag_{i\sigma}c_{i\sigma}$. We consider a square lattice with 
periodic-antiperiodic boundary conditions and choose $U$ to be equal to the 
bandwidth, $U=8t$. Our ansatz 
\begin{equation} 
\qquad\vert\Psi\rangle=e^{-h\hat T}e^{-g\hat D}\vert\Psi_{0}\rangle\, 
\end{equation}
is linked to a mean-field ground state $\vert\Psi_{0}\rangle$ with either a 
($d$-wave) superconducting or an antiferromagnetic order parameter. The
operator $e^{-g\hat D}$ partially suppresses double occupancy for $g>0$, while
$e^{-h\hat T}$ promotes both hole motion and kinetic exchange 
(close to half filling). In the limit $h\rightarrow 0$ we recover the 
Gutzwiller ansatz \cite{Giam1}. For $g\rightarrow\infty$ and $h\ll 1$ our 
variational problem is equivalent to that of the $t$-$J$ Hamiltonian with 
respect to a fully Gutzwiller-projected mean-field state. 

The calculations for $h>0$ are carried out in momentum space 
where the operator $\hat D$ 
is not diagonal. Therefore a discrete Hubbard-Stratonovich transformation 
is applied to decouple the terms $n_{i\uparrow}n_{i\downarrow}$ in the 
operator $e^{-g\hat{D}}$ by introducing an Ising spin at 
each site. Expectation values are obtained using a 
Monte Carlo simulation with respect to the Ising spin configurations.

We discuss first the case of an average site occupation $n=1$
(half-filling), where the ground state is expected to exhibit long-range
antiferromagnetic order. A commensurate spin-density wave characterized by
a gap parameter $\Delta_{AF}$ is therefore the natural mean-field reference
state, $\vert\Psi_{0}\rangle=\vert\mbox{SDW}\rangle$.
The order parameter is the staggered magnetization defined by
\begin{equation}
M=\frac{1}{N}\sum_{i}(-1)^{i}\langle n_{i\uparrow}-n_{i\downarrow}\rangle\, ,
\end{equation}
where N is the number of sites. In Table \ref{half_tab} the results obtained 
by minimizing the energy expectation value with respect to the three 
variational parameters $g,h, \Delta_{AF}$ for an $8\times 8$ lattice are 
compared with the unrestricted Hartree-Fock approximation ($g=h=0$), the 
Gutzwiller wave function ($g>0, h=0$)
\cite{Yoko3}, a quantum Monte Carlo simulation 
\cite{Hirs1} and a Projector Operator technique \cite{Pola1}. The gap 
parameter $\Delta_{AF}$ is very large for $g=h=0$ and decreases dramatically 
if $g$ and $h$ are optimized. We note that 
the gap parameter cannot be identified with an excitation gap, which 
in fact should vanish if a continuous symmetry is broken. The ground state 
energy is seen to vary appreciably as the parameters $g$ and $h$ are turned 
on and to be comparable to that found with other techniques. On the other 
hand, the order parameter is still rather large, at least in comparison to
the accepted value of $M=0.614(1)$ for the 2D Heisenberg model ($U=\infty$)
\cite{Sand1}, an upper bound for the Hubbard model.  

\begin{center}
\begin{table}[!h]
\vspace{-0.2cm}
\begin{tabular}{lccccc}
\hline
\hline
&g&h&$\Delta_{AF}$&$M$&E/t\\
\hline
VMC&0&0&3.6(1)&0.89(1)&-0.466(1)\\
VMC&0.69&0&1.3&0.86(1)&-0.493(3)\\
VMC&3.1(1)&0.101(3)&0.32(2)&0.77(1)&-0.514(1)\\
QMC&-&-&-&0.42(1)&-0.48(5)\\
PO&-&-&-&-&-0.521(1)\\
\hline
\hline
\end{tabular}
\caption{\label{half_tab} Variational results (VMC) for the 2D Hubbard model 
at half-filling ($8\times 8$ lattice, $U=8t$), compared to quantum Monte Carlo 
simulations (QMC) and a Projector Operator approach (PO). The VMC data 
include the unrestricted Hartree-Fock 
approximation, the Gutzwiller ansatz and the present study.} 
\end{table}
\end{center}

\vspace{-1cm}
In order to extract some information about superconducting correlations
in the presence of antiferromagnetic long-range order, we have calculated the 
correlation function $F_{ij}=\langle C_{i}^{\dag}C_{j}\rangle$, where
\begin{equation}
C_{i}^{\dag}
=\sum_{j_{i}}\sigma_{j_{i}}(c_{i\uparrow}^{\dag}c_{j_{i}\downarrow}^{\dag}
-c_{i\downarrow}^{\dag}c_{j_{i}\uparrow}^{\dag})\ .
\end{equation}
The sites $j_{i}$ are the four nearest neighbors of site $i$ and
$\sigma_{j_{i}}=+1(-1)$ in $x$-($y$-) direction. Thus  $C_{i}^{\dag}$ creates 
a singlet pair with {\it d}-wave symmetry centered at site $i$. $F_{ij}$ is 
found to decrease rapidly with increasing distance, as 
expected for a gapped system. For on-site correlations we find 
$F_{ii}=0.0637(1)$ for $h\ne 0$ and $F_{ii}=0.0592(1)$ for $h=0$, while the
results for nearest-neighbor correlations are $F_{ij}=0.0171(1)$ for 
$h\ne 0$ and $F_{ij}=0.0155(1)$ for $h=0$. The superconducting correlations 
are therefore slightly enhanced by the parameter $h$.

We now discuss the effects of hole doping, and in particular the possibility 
of $d$-wave superconductivity, as suggested by Renormalization Group 
arguments \cite{Zanc1,Halb1}, previous variational calculations
\cite{Gros1,Yoko1,Para1,Giam1,Sore1,Yoko2} and quantum Monte Carlo 
simulations \cite{Maie1}.  Our mean-field reference state is the BCS 
wave function with {\it d}-wave symmetry, 
$\vert\Psi_{0}\rangle=\vert d\mbox{BCS}\rangle$, characterized by a gap 
parameter $\Delta$ describing pairing and a ``chemical potential'' $\mu$ 
fixing the average electron density $n$ \cite{Eich1}. 
To reduce the 
statistical error in the Monte Carlo simulations and consequently the 
computational time, a fixed set of ``Ising spin'' configurations is first 
generated and then used to optimize the variational parameters 
\cite{Kalo1,Umri1}. 

\begin{center}
\begin{table}[!t]
\begin{tabular}{ccccc}
\hline
\hline
n&\hspace{0.4cm}$\mu$\hspace{0.4cm}&\hspace{0.4cm}g\hspace{0.4cm}&\hspace{0.4cm}h\hspace{0.4cm}&\hspace{0.4cm}E/t\hspace{0.4cm}\\
\hline
0.8125&\hspace{0.4cm}-0.4418(1)\hspace{0.4cm}&\hspace{0.4cm}3.0(1)\hspace{0.4cm}&\hspace{0.4cm}0.099(2)\hspace{0.4cm}&\hspace{0.4cm}-0.849(1)\hspace{0.4cm}\\
0.8400&\hspace{0.4cm}-0.3972(3)\hspace{0.4cm}&\hspace{0.4cm}3.2(1)\hspace{0.4cm}&\hspace{0.4cm}0.103(2)\hspace{0.4cm}&\hspace{0.4cm}-0.802(1)\hspace{0.4cm}\\
0.9000&\hspace{0.4cm}0.357(1)\hspace{0.4cm}&\hspace{0.4cm}3.4(1)\hspace{0.4cm}&\hspace{0.4cm}0.106(2)\hspace{0.4cm}&\hspace{0.4cm}-0.697(1)\hspace{0.4cm}\\
0.9375&\hspace{0.4cm}0.620(1)\hspace{0.4cm}&\hspace{0.4cm}3.7(1)\hspace{0.4cm}&\hspace{0.4cm}0.110(2)\hspace{0.4cm}&\hspace{0.4cm}-0.627(1)\hspace{0.4cm}\\
0.9600&\hspace{0.4cm}0.692(1)\hspace{0.4cm}&\hspace{0.4cm}4.1(2)\hspace{0.4cm}&\hspace{0.4cm}0.115(3)\hspace{0.4cm}&\hspace{0.4cm}-0.583(1)\hspace{0.4cm}\\
0.9700&\hspace{0.4cm}0.743(1)\hspace{0.4cm}&\hspace{0.4cm}4.3(2)\hspace{0.4cm}&\hspace{0.4cm}0.116(3)\hspace{0.4cm}&\hspace{0.4cm}-0.564(1)\hspace{0.4cm}\\
\hline
\hline
\end{tabular}
\caption{\label{doped_tab}``Chemical potential'', parameters g and h and total energy per site for different densities on an 8x8 lattice.}
\end{table}
\end{center}

\vspace{-1cm}
The ground state energy and the parameters $g, h, \mu$
are given in Table \ref{doped_tab} for an 8x8 lattice and various densities.
The chemical potential $\mu$ varies strongly with doping and increases so much 
for $n\rightarrow 1$ that the optimization becomes very difficult. The 
Gutzwiller parameter $g$ also increases rather strongly for $n\rightarrow 1$,
which indicates that the system is ``more localized'' at half-filling than 
away from half-filling \cite{Baer1}. In contrast, the kinetic parameter $h$
does not vary appreciably.

\begin{figure}[!t]
\includegraphics[width=0.45\textwidth]{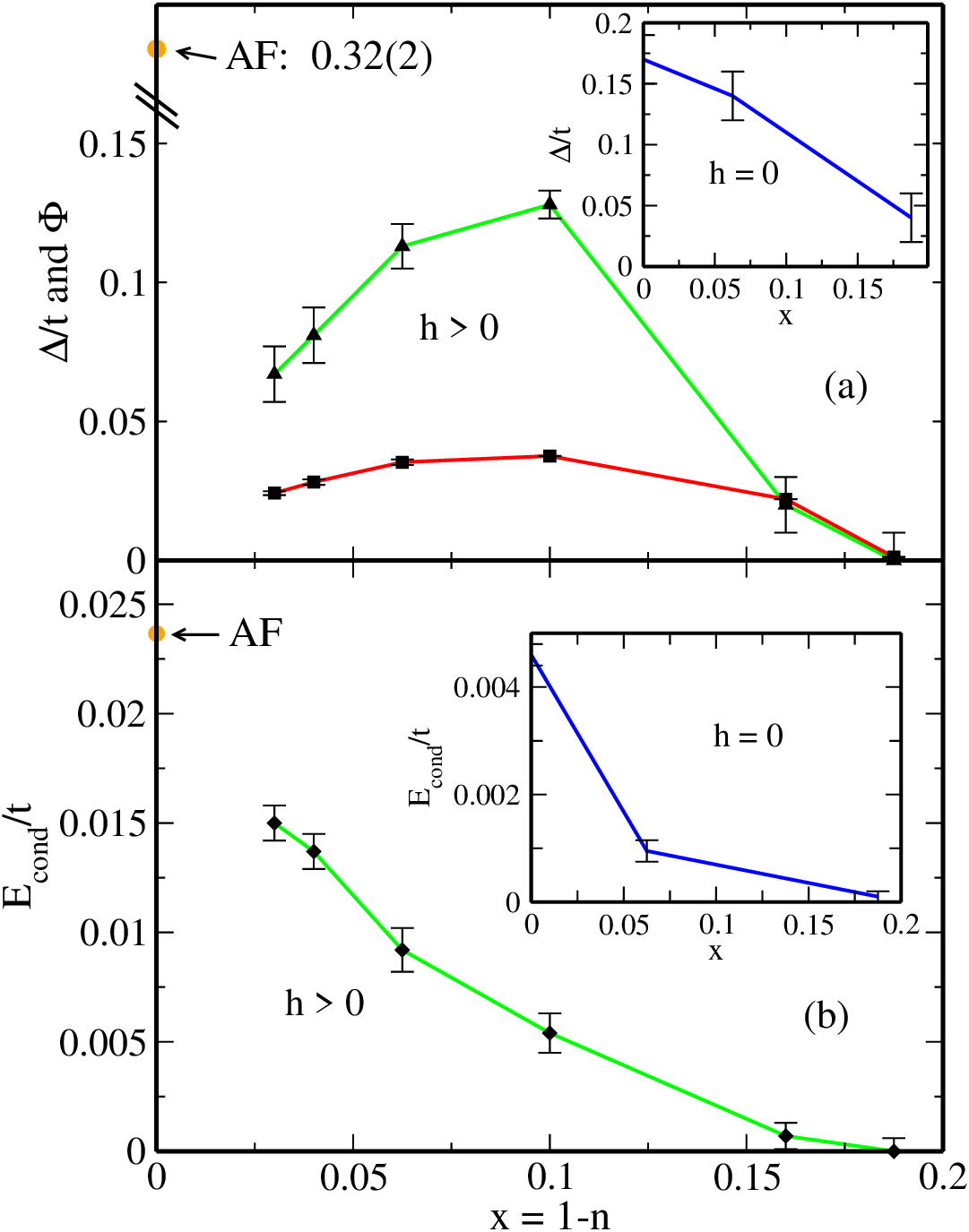}
\caption{\label{D+E}(Color online) (a): Gap (triangles) and order parameter 
(squares) as functions of the doping for an 8x8 lattice. 
The mark at half-filling is the antiferromagnetic gap. 
The inset shows the gap parameter for the Gutzwiller wave function. 
(b): Condensation energy per site. (a)+(b): Error bars indicate statistical 
uncertainties.}
\end{figure}

The gap parameter $\Delta$ and the order parameter 
$\Phi=\vert\langle c_{i\uparrow}^{\dag}c_{j_{i}\downarrow}^{\dag}\rangle\vert$
are shown in Fig.\ \ref{D+E}(a) as functions of the hole density
$x=1-n$, for an $8\times 8$ lattice. Both quantities have a maximum around
$x=0.1$ and tend to zero around $x=0.18$. The limiting behavior for 
$x\rightarrow 0$ has not been established firmly, due to computational 
problems mentioned above, but our results are consistent with
$\Delta\rightarrow 0,\ \Phi\rightarrow 0$. For the Gutzwiller wave function,
the order parameter also exhibits a dome shape, but not so the gap parameter:
$\Delta$ is found to increase monotonically for $x\rightarrow 0$,
both for finite $U$ (inset of Fig.\ \ref{D+E}(a)) and for $U\rightarrow\infty$ 
\cite{Para1}.

The condensation energy, $E_{cond}=E(0)-E(\Delta)$ where $\Delta$ is the 
optimal gap parameter, is depicted in Fig.\ \ref{D+E}(b). It vanishes for
$x>0.18$ and increases monotonically with decreasing $x$, even beyond the hole
concentration where both $\Delta$ and $\Phi$ pass through a maximum. The
limiting behaviour for $x\rightarrow 0$ is again unknown, but for $x=0$
antiferromagnetism prevails. The comparison with the Gutzwiller 
wave function (inset) indicates that the addition of the parameter $h$ strongly
enhances the condensation energy \cite{Cond1}. It is worthwhile to
add that according to calculations for small clusters \cite{Otsu1} the
difference $\Delta E= E_{var}-E_0$ between the variational energy $E_{var}$
and the exact ground state energy $E_0$ is of the same order for $h>0$
as the condensation energy $E_{cond}$ ($\Delta E \approx 0.007t$,
$E_{cond}\approx 0.005t$ at $n=0.9$), in contrast to the 
case $h=0$ where $\Delta E\gg E_{cond}$ ($\Delta E \approx 0.08t,\ 
E_{cond}\approx 0.001t$).

\begin{figure}[!t]
\vspace{0.5cm}
\includegraphics[width=0.45\textwidth]{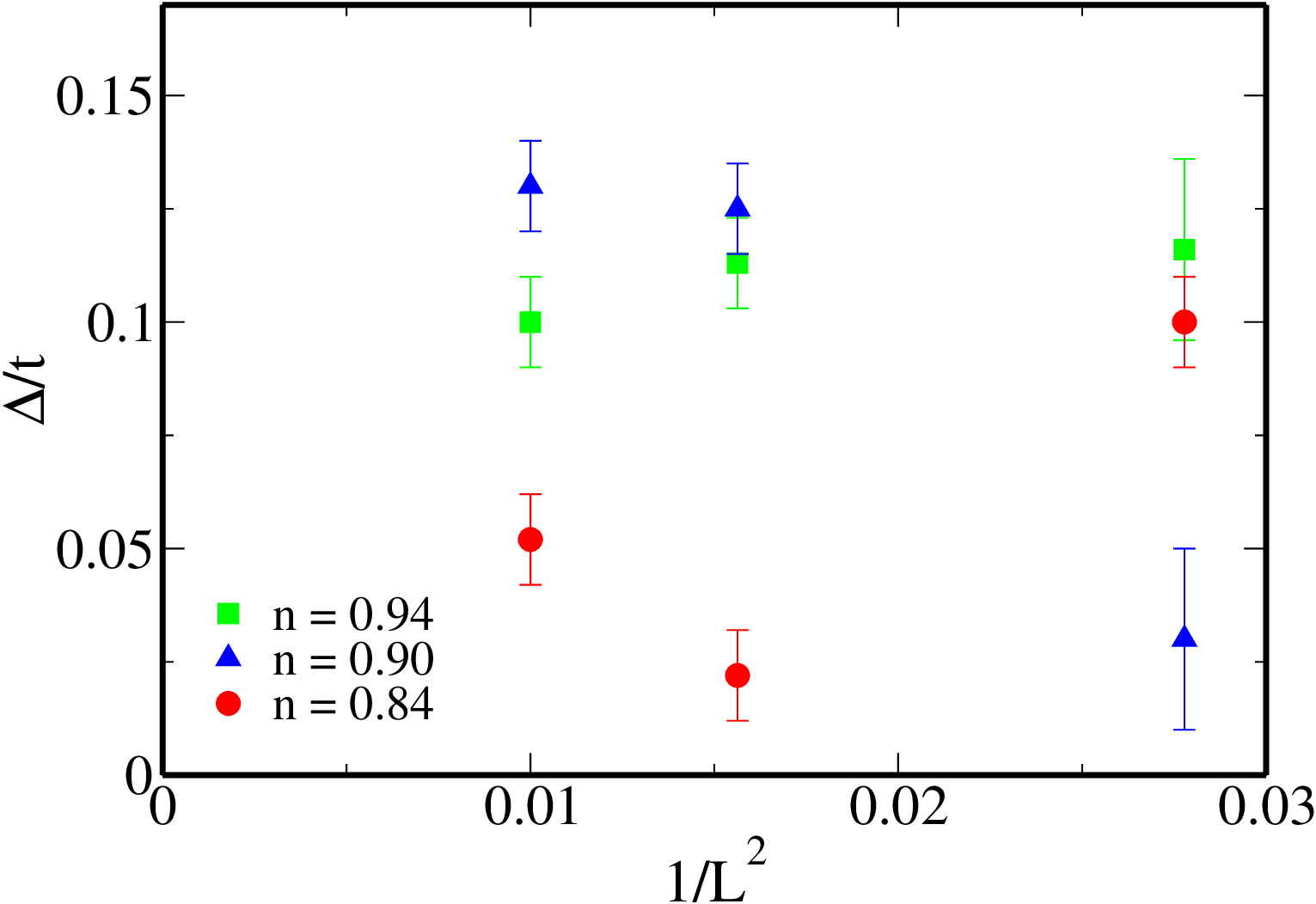}
\caption{\label{size}(Color online) Finite-size scaling of the gap parameter for densities $n=0.84$ (circles), $n=0.90$ (triangles) and $n=0.9375$ (squares).}
\end{figure}

An important question is to what extent an $8\times 8$ lattice is able to
mimic the thermodynamic limit. Therefore we have also studied other lattice 
sizes. The results for the gap parameter (Fig.\ \ref{size}) show that the size 
effects are more important in regions where the gap is small ($n=0.84$) 
than well inside the superconducting dome ($n=0.90$ or $n=0.9375$). However, 
even at $n=0.90$ where the gap is maximal, a $6\times6$ lattice is not large 
enough to give a reliable estimate for the thermodynamic limit.

\begin{figure}[!t]
\includegraphics[width=0.45\textwidth]{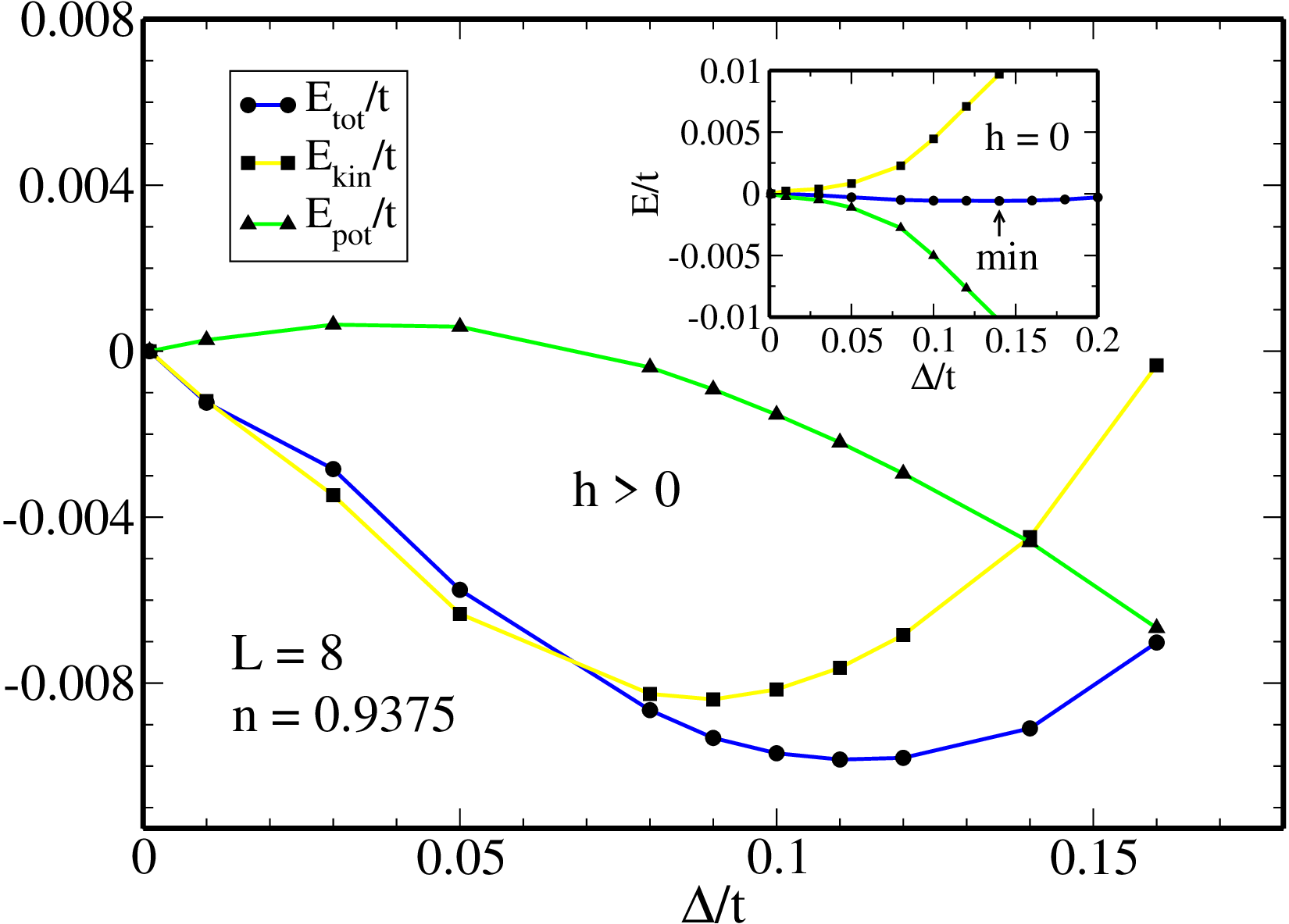}
\caption{\label{energies}(Color online) Total (circles), kinetic (squares) and 
potential (triangles) energies per site as functions of the gap parameter on 
an 8x8 lattice, for the density $n=0.9375$. For each curve $E(\Delta=0)$ has 
been subtracted. The relative error is smaller than the symbol size. The 
corresponding results for the Gutzwiller wave function are given in the inset.}
\end{figure}

In Fig.\ \ref{energies} the kinetic and the potential energies are plotted 
separately for a density $n=0.9375$ as functions of the gap parameter. 
It turns out that the maximum energy gain 
(the condensation energy) at $\Delta=0.11t$ is to a large extent ($>75\%$) due 
to a decrease in the kinetic energy, in contrast to the BCS behaviour where 
the condensation energy is entirely due to the potential energy. Our findings
are also qualitatively different from those obtained with a Gutzwiller wave 
function for which the kinetic energy increases monotonically with the gap 
parameter (inset). 
We have also studied the magnetic structure factor
\begin{equation}
S({\bf q})=\frac{1}{N}\sum_{i,j}e^{i{\bf q}\cdot 
({\bf R_{i}}-{\bf R_{j}})}\langle(n_{i\uparrow}-n_{i\downarrow})
(n_{j\uparrow}-n_{j\downarrow})\rangle\ .
\end{equation}
within the superconducting phase.
\begin{figure}[!t]
\includegraphics[width=0.45\textwidth]{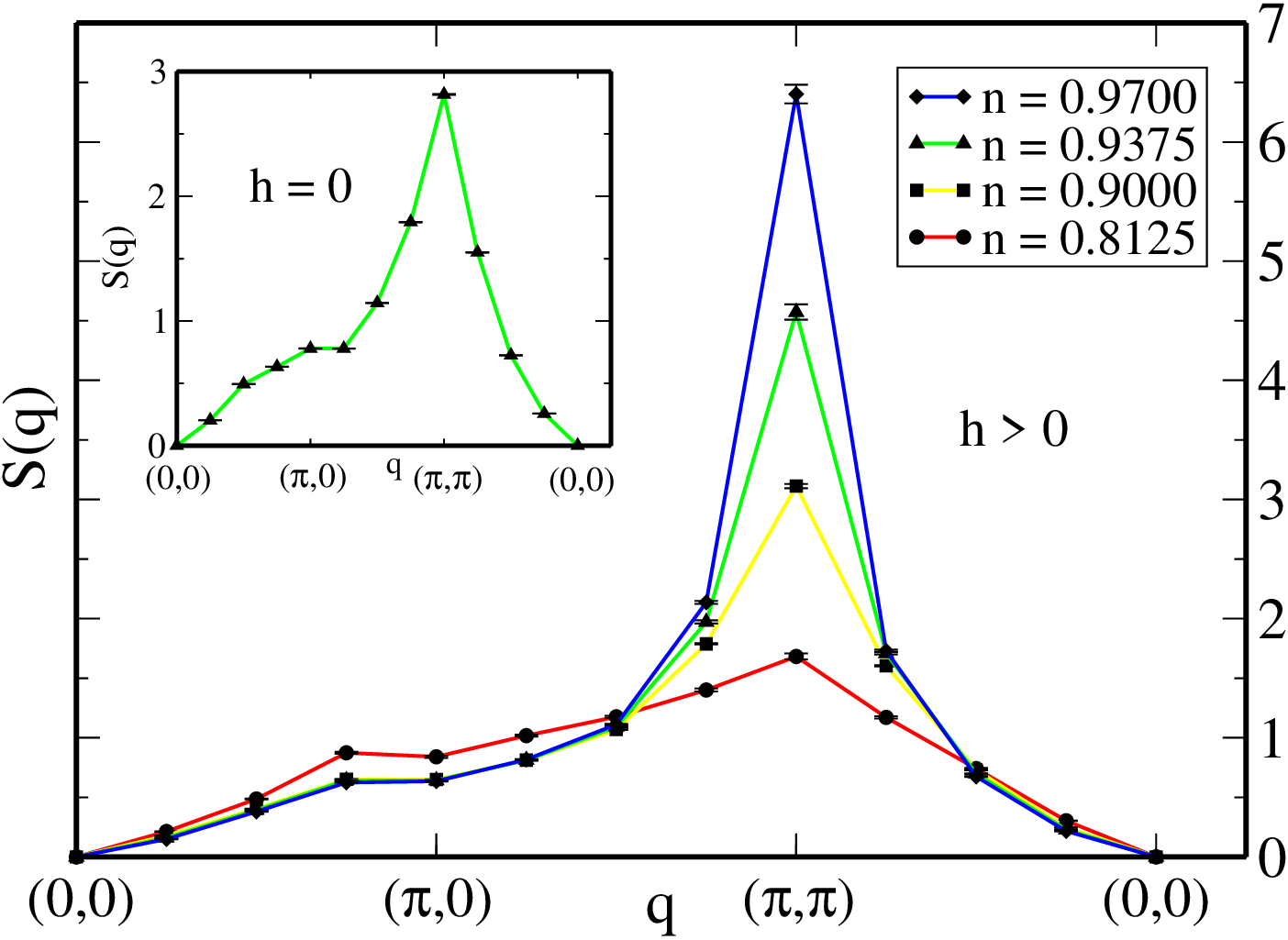}
\caption{\label{sp_cor}(Color online) Magnetic structure factor as a function 
of the wave vector for different densities and an 8x8 lattice. The inset 
shows the magnetic structure factor for the Gutzwiller wave function 
at $n=0.9375$.}
\end{figure}
Fig.\ \ref{sp_cor} shows this function for several densities along three 
different lines in the Brillouin zone. The structure factor is peaked 
at $(\pi,\pi)$, indicating antiferromagnetic correlations. The peak decreases 
with increasing  hole concentration. The comparison with results for
$h=0$ (inset) shows that the antiferromagnetic correlations are strongly 
enhanced by the parameter $h$.

In summary, we have found that the addition of a ``kinetic projector'' 
$e^{-h\hat{T}}$ to the Gutzwiller wave function 
yields both quantitative improvements (for instance for the ground state 
energy or for the antiferromagnetic order parameter) and qualitative changes
(such as the doping dependence of the superconducting gap or the decrease of 
the kinetic energy as a function of the gap parameter). Nevertheless, there 
remains room for improvement because our variational ansatz (as well as
all trial states used previously) is linked to a delocalized mean-field
reference state and thus requires a strong suppression of double occupancy,
at least for $U=8t$. This reflects the fact that this
parameter regime corresponds to that of a doped Mott insulator which would
be treated more naturally starting from a localized reference state.

Finally, we comment on the relevance of our findings for layered 
cuprates. The antiferromagnetic ground state for $x=0$ and a superconducting 
phase with $d$-wave symmetry are well established experimentally, with a 
slightly different doping range for superconductivity ($0.05<x<0.3$ against 
$0<x<0.18$ in our study). The addition of a hopping term between next-nearest
neighbor sites (parameter $t'$) might improve this comparison. 
The typical size
of the superconducting gap at optimal doping is 30 meV \cite{Dama1}, in good agreement with our result 
($\Delta\approx 0.13t \approx 40$ meV for $t=300$ meV). The comparison with
condensation energies extracted from specific heat measurements is less 
encouraging ($0.10-0.20$ meV/copper \cite{Lora1,vanH1} against 
$0.005t=1.5$ meV in this work). 
Much discussion has been raised by the question of ``kinetic energy driven
superconductivity'' where, in contrast to BCS, the energy gain arises from
a decrease in kinetic energy. The reported gain of kinetic energy 
$\Delta E_{kin}\approx 0.5-1.0$ meV on the basis of optical spectroscopy 
at optimal doping \cite{vanH1,Mol1} 
corresponds well to our result ($\Delta E_{kin}\approx 1.1$ meV), but we
have to be aware that the use of a low-energy cut-off in the 
frequency-integration of the conductivity is not unambiguous. 
Good agreement with experiment is also found for the strong antiferromagnetic
correlations in the superconducting phase. In fact, the structure factor
$S({\bf q})$ determined by neutron scattering experiments shows a pronounced
peak at $(\pi,\pi)$, which decreases upon doping \cite{Fong1}.

In conclusion, the present variational calculations give an 
appealing picture of the ground state of the 2D Hubbard model, both at 
half-filling and for the doped system. Superconductivity out of purely
repulsive interactions appears very naturally in this scheme. Several
predictions agree surprisingly well with experiments on layered cuprates.  

We are grateful for financial support from the Swiss National Science Foundation through the National Center of Competence in Research ``Materials with Novel Electronic Properties-MaNEP''.


\begin{thebibliography}{25}
\bibitem{Cond1} It should be mentioned that the
calculations for $h>0$ have been carried out for a ``grand canonical ensemble''
(where $\mu$ is fixed by the average density), whereas the results for
$h=0$ have been obtained for a ``canonical ensemble'' (where the particle 
number is fixed and $\mu$ is a variational parameter). 
The two ensembles are equivalent in the thermodynamic limit.
\end{thebibliography}
\end{document}